\documentclass[sigconf]{acmart}

\usepackage{multirow}
\usepackage{booktabs}
\usepackage{graphicx}
\usepackage{subfigure}
\usepackage{float}
\usepackage{svg}
\usepackage{tcolorbox}
\usepackage{enumitem}

\usepackage{footmisc}
\newcommand{\LossName}{BCE loss}

\AtBeginDocument{%
  \providecommand\BibTeX{{%
    \normalfont B\kern-0.5em{\scshape i\kern-0.25em b}\kern-0.8em\TeX}}}
\setcopyright{acmcopyright}
\copyrightyear{2024}
\acmYear{2024}
\setcopyright{rightsretained}
\acmConference[KDD '24]{Proceedings of the 30th ACM SIGKDD Conference on Knowledge Discovery and Data Mining}{August 25--29, 2024}{Barcelona, Spain}
\acmBooktitle{Proceedings of the 30th ACM SIGKDD Conference on Knowledge Discovery and Data Mining (KDD '24), August 25--29, 2024, Barcelona, Spain}\acmDOI{10.1145/3637528.3671565}
\acmISBN{979-8-4007-0490-1/24/08}
\begin{document}
\pagenumbering{arabic}
\title{Understanding the Ranking Loss for Recommendation with Sparse User Feedback}
% TODO
% Grouping authors’ names or e-mail addresses, or providing an
% “e-mail alias,” as shown below, is not acceptable:
\author{Zhutian Lin}
\authornote{Both authors contributed equally to this research}
\affiliation{
  \institution{Shenzhen International Graduate School, Tsinghua University}
  % \city{Shenzhen}
  \country{}
}
\email{linzt22@mails.tsinghua.edu.cn}

\author{Junwei Pan}
\authornotemark[1]
\affiliation{
  \institution{Tencent Inc.}
  \country{}
  % \city{Shenzhen}
  % \country{China}
}
\email{jonaspan@tencent.com}

\author{Shangyu Zhang}
\affiliation{
  \institution{Tencent Inc.}
  \country{}
  % \city{Shenzhen}
  % \country{China}
}
\email{vitosyzhang@tencent.com}

\author{Ximei Wang}
\affiliation{
  \institution{Tencent Inc.}
  \country{}
  % \city{Shenzhen}
  % \country{China}
}
\email{messixmwang@tencent.com}

\author{Xi Xiao}
\authornote{Corresponding Author}
\affiliation{
  \institution{Shenzhen International Graduate School, Tsinghua University}
  \country{}
  % \city{Shenzhen}
  % \country{China}
}
\email{xiaox@sz.tsinghua.edu.cn}

\author{Shudong Huang}
\affiliation{
  \institution{Tencent Inc.}
  \country{}
  % \city{Shenzhen}
  % \country{China}
}
\email{ericdhuang@tencent.com}

\author{Lei Xiao}
\affiliation{
  \institution{Tencent Inc.}
  \country{}
  % \city{Shenzhen}
  % \country{China}
}
\email{shawnxiao@tencent.com}

\author{Jie Jiang}
\affiliation{
  \institution{Tencent Inc.}
  \country{}
  % \city{Shenzhen}
  % \country{China}
}
\email{zeus@tencent.com}

% \author{Zhutian Lin$^{1*}$, Junwei Pan$^{2*\dagger}$, Shangyu Zhang$^2$, Ximei Wang$^2$\\
% Xi Xiao$^{1\dagger}$, Shudong Huang$^2$, Lei Xiao$^2$, Jie Jiang$^2$}
% \affiliation{%
%     \institution{$^1$ Shenzhen International Graduate School, Tsinghua University \\
%     $^2$ Tencent
%     \country{}
% }}
% \email{{linzt22}@mails.tsinghua.edu.cn,
% {jonaspan, vitosyzhang, messixmwang}@tencent.com}
% \email{{xiaox}@sz.tsinghua.edu.cn,
% {ericdhuang, shawnxiao, zeus}@tencent.com}

\renewcommand{\shortauthors}{Lin Z. and Pan J. et al.}

\begin{abstract}

% Click-through rate (CTR) prediction is a critical research topic in online advertising. 
% While many existing approaches treat it as a binary classification problem and utilize binary cross entropy (BCE) as the optimization objective, recent advancements have indicated that combining BCE loss with an auxiliary ranking loss yields substantial performance improvements. 
% However, the full efficacy of this combination loss remains incompletely understood.
% In this paper, we discover a new challenge associated with BCE loss in scenarios with sparse positive feedback: \emph{the gradient vanishing of negative samples}. 
% We then introduce a novel perspective on the effectiveness of the auxiliary ranking loss in CTR prediction: it \emph{generates larger gradients on negative samples}, thereby \emph{mitigating the optimization difficulties when using the BCE loss only} and resulting in \emph{improved classification ability}.
% We validate this perspective through theoretical analysis and extensive empirical evaluation on public datasets. 
% Furthermore, we successfully introduced the ranking loss in Tencent's online advertising system, achieving notable lifts of 0.70\% and 1.26\% in Gross Merchandise Value (GMV) for two main scenarios. 
% The code for our approach is openly accessible at: \url{https://github.com/SkylerLinn/Understanding-the-Ranking-Loss}.

Click-through rate (CTR) prediction is a crucial area of research in online advertising.
While binary cross entropy (BCE) has been widely used as the optimization objective for treating CTR prediction as a binary classification problem, recent advancements have shown that combining BCE loss with an auxiliary ranking loss can significantly improve performance. 
However, the full effectiveness of this combination loss is not yet fully understood.
In this paper, we uncover a new challenge associated with the BCE loss in scenarios where positive feedback is sparse: \emph{the issue of gradient vanishing for negative samples}. 
We introduce a novel perspective on the effectiveness of the auxiliary ranking loss in CTR prediction: it \emph{generates larger gradients on negative samples}, thereby \emph{mitigating the optimization difficulties when using the BCE loss only} and resulting in \emph{improved classification ability}.
To validate our perspective, we conduct theoretical analysis and extensive empirical evaluations on public datasets. 
Additionally, we successfully integrate the ranking loss into Tencent's online advertising system, achieving notable lifts of 0.70\% and 1.26\% in Gross Merchandise Value (GMV) for two main scenarios. 
The code is openly accessible at: \url{https://github.com/SkylerLinn/Understanding-the-Ranking-Loss}.
\end{abstract}

\begin{CCSXML}
<ccs2012>
   <concept>
       <concept_id>10002951.10003260.10003272.10003275</concept_id>
       <concept_desc>Information systems~Display advertising</concept_desc>
       <concept_significance>500</concept_significance>
       </concept>
   <concept>
       <concept_id>10010147.10010257.10010293.10010294</concept_id>
       <concept_desc>Computing methodologies~Neural networks</concept_desc>
       <concept_significance>500</concept_significance>
       </concept>
   <concept>
       <concept_id>10010147.10010257.10010293.10010309</concept_id>
       <concept_desc>Computing methodologies~Factorization methods</concept_desc>
       <concept_significance>500</concept_significance>
       </concept>
 </ccs2012>
\end{CCSXML}

\ccsdesc[500]{Information systems~Display advertising}
\ccsdesc[500]{Computing methodologies~Neural networks}
\ccsdesc[500]{Computing methodologies~Factorization methods}

\keywords{Recommendation Systems; CTR Prediction; Gradient Vanishing; Ranking Loss}

\maketitle
% \renewcommand{\thefootnote}{\fnsymbol{footnote}}
% \footnotetext[1]{Equal contribution}
% \footnotetext[2]{Corresponding author}
% \renewcommand{\thefootnote}{\arabic{footnote}}

\section{Introduction}

In recent decades, users have encountered abundant information while navigating websites or mobile applications. 
This inundation presents significant challenges for electronic retailers, content providers, and online advertising platforms as they strive to recommend appropriate items to individual users within specific contexts. 
Thus, the deployment of recommendation systems has become widespread, enabling the prediction of users' preferences from a vast pool of candidate items.

For instance, in effective cost per mille (eCPM) advertising, advertising platforms must bid for each advertisement based on the estimated value of the impression, which relies on the bid value and the estimated Click-through rate (CTR). 
Consequently, accurately predicting user response has emerged as a critical factor, attracting substantial research attention.

CTR prediction~\cite{chen2020simple, trench2013, fm, he2014practical} is usually formulated as a binary classification problem and optimized by a binary cross entropy (BCE) loss.
Some recent works~\cite{twitter, yan2022scale, jrc, bai2023regression} in the industry propose to combine the BCE loss with an auxiliary \emph{ranking loss}, which was usually used in Learning to Rank (LTR)~\cite{ranknet, burges2006learning, cao2007learning, zheng2007regression, liu2009learning}.
For example, Combined-Pair~\cite{twitter} stands as one of the pioneering attempts to combine a pairwise loss with a pointwise loss for CTR prediction in the Twitter Timeline.
\citeauthor{yan2022scale} and ~\citeauthor{bai2023regression} proposed to combine regression loss with ranking loss to better trade-off between the regression and ranking objectives~\cite{yan2022scale, bai2023regression}.
~\citeauthor{jrc} proposed to employ two logits to optimize ranking and calibration objectives~\cite{jrc} jointly.
Such combination loss is widely adopted in real-world recommendation systems in Twitter~\cite{twitter}, Google~\cite{yan2022scale, bai2023regression}, and Alibaba~\cite{jrc}.

The prevailing literature predominantly attributes the success of combining classification and ranking losses to the augmentation of ranking capability~\cite{twitter, jrc}.
These studies substantiate their hypothesis by observing an enhancement in the Area Under the Curve (AUC), a metric commonly employed to evaluate ranking quality.
However, our curiosity lies in investigating the impact of the combination loss on the model's primary optimization objective: the classification ability, as measured by the \LossName{} metric. 
To explore this, we conducted a comparative analysis between the BCE method, which solely employs the Binary Cross-Entropy (BCE) loss, and the Combined-Pair approach, which incorporates a BCE-ranking combination loss. 
Our evaluation was performed on the Criteo dataset, with artificial weights on positive samples to simulate the sparse positive scenario.

\begin{figure}[!tp]
    \centering
    \subfigure[BCE Loss Dynamics along epochs.] {
        \includegraphics[width=.25\textwidth]{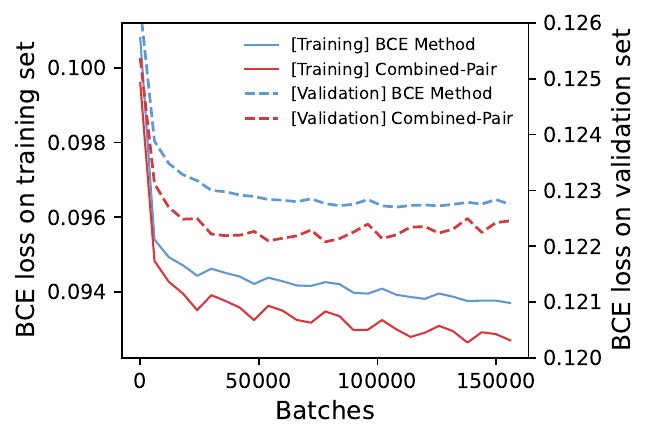}
        \label{fig: finding_a}
    }
    \subfigure[Loss Landscapes.] {
        \centering\includegraphics[width=.20\textwidth]{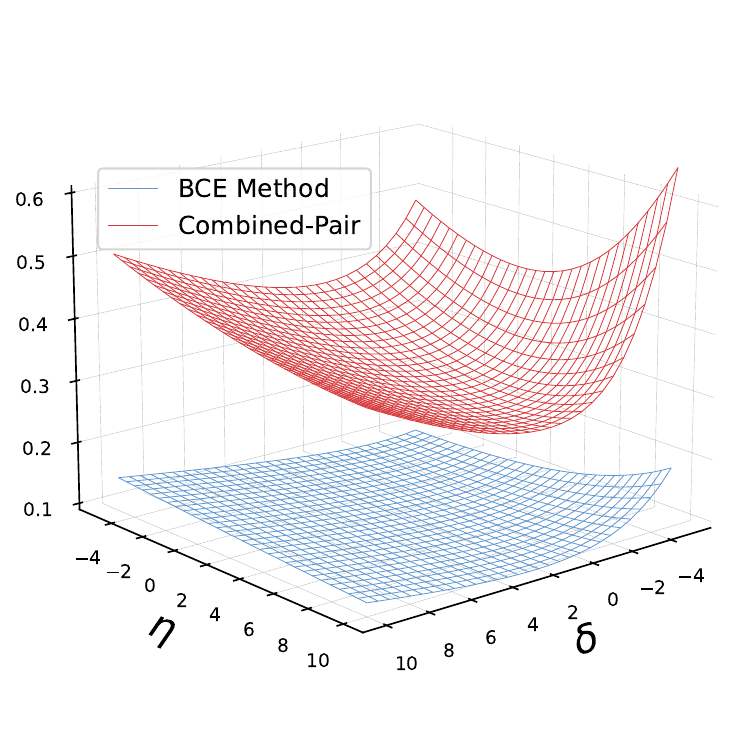}
        \label{fig: finding_c}
    }
    \caption{
        (a) BCE Loss Dynamics along epochs on the training and validation set and (b) Loss Landscape of BCE method (blue) and Combined-Pair (red).
    }
\end{figure}

Surprisingly, our findings on the validation set, as depicted in Fig.~\ref{fig: finding_a}, revealed \emph{a reduction in the \LossName{} of the Combined-Pair method (the red dashed line) compared to that of the BCE method (the blue dashed line)}. 
This intriguing observation suggests that \emph{the inclusion of an additional ranking loss not only enhances the model's ranking ability but also improves its classification ability}.

To investigate the cause of this improvement further, we delve into the optimization procedure during model training.
% two potential factors: the model's \emph{generalization capability} and \emph{optimization procedure during training}. 
In particular, we monitor the \LossName{} of these two methods during model training, as the solid lines in Fig.~\ref{fig: finding_a}. 
To our surprise, we find that \emph{the BCE loss of the Combined-Pair (the red solid line) experiences a significant reduction compared to that of the BCE method (the blue solid line)}.
Moreover, we visualize the loss landscape of these two methods using Contour Plots \& Random Directions~\cite{goodfellow2014qualitatively}.
Our analysis reveals that the BCE method exhibits a relatively flat landscape, indicating a slower optimization process. In contrast, the Combined-Pair method demonstrates a significantly steeper landscape, as illustrated in Fig.~\ref{fig: finding_c}.

% This observation suggests that the BCE method is susceptible to gradient vanishing issues.

We proceed to investigate the gradients of the BCE method and the Combined-Pair. 
In scenarios with sparse positive feedback, such as ad CTR prediction where only a small fraction of samples are positive (clicks), we demonstrate that \emph{negative samples get small gradients in the BCE method, leading to optimization difficulties}. 
However, \emph{the ranking loss in the Combined-Pair contributes significantly larger gradients, effectively mitigating the gradient vanishing problem}.

To further validate our perspective, we conduct comprehensive experiments. 
Firstly, we generate artificial datasets with varying degrees of positive sample sparsity and observe that the sparser the positive samples, the greater the performance improvement achieved by incorporating an auxiliary ranking loss. 
Secondly, in addition to the classification-ranking combination loss, we explore alternative approaches that address the gradient vanishing issue, such as Focal Loss~\cite{lin2017focal} and a new Combined-Contrastive method. 
Lastly, we successfully deployed the classification-ranking combination loss in the CTR prediction within Tencent's online advertising system, resulting in substantial revenue increases.
In summary, our contributions can be summarized as follows:
\begin{itemize}
    \item We uncover a challenge associated with binary cross entropy loss in recommendation scenarios with sparse positive feedback: the gradient vanishing of negative samples.
    \item We present a novel perspective on the effectiveness of involving an auxiliary ranking loss in recommendation systems: it introduces larger gradients for negative samples, addressing the gradient vanishing issue.
    \item We substantiate our claims through theoretical analysis, offline experiments, and online empirical evaluations. 
\end{itemize}

\section{Related Work}

\subsection{Click-Through Rate Prediction}

In online advertising, the main objective of CTR prediction is to estimate the likelihood of a user $u\in U$ clicking on a given ad $i\in I$. 
The input $(x, y)\sim(X, Y)$ represents an impression of an ad to a specific user, where $x$ represents the features of that request, and $y\in \{0,1\}$ serves as the click label. 
For any given model $f_\theta:\mathcal{R}^d\rightarrow \mathcal{R}$, parameterized by $\theta$, the logit is obtained by $z = f_\theta(x)$. 
Commonly, it is passed through a sigmoid function $\sigma(\cdot)$ to obtain the estimated value as the output, which can be expressed as $\hat{p}=p(y=1|x)=\sigma(z)$.

Current research in the CTR prediction task primarily focuses on model structure and optimization objectives. 
Regarding model structure, recent studies have primarily aimed to explore ways to capture higher-order interaction information effectively. 
This has led to the proposal of various model structures such as Factorization Machines (FM) family \cite{rendle2010factorization, ffm2017, fwfm2018, fmfm2021}, Wide\&Deep~\cite{wideanddeep2017}, DeepFM~\cite{deepfm2017}, IPNN~\cite{pnn2016}, xDeepFM~\cite{xdeepfm2018}, DCN V2 \cite{wang2021dcn}, Final MLP~\cite{mao2023finalmlp} and Multi-Embedding~\cite{me2023}. 
As for optimization objectives, three main loss paradigms have been introduced: pointwise, pairwise, and listwise approaches, which will be discussed in the following sections.

\subsection{Binary Cross Entropy}
In a pointwise approach, each item is treated \textit{independently}, and the objective is to optimize each item's prediction or relevance score directly. 
One commonly used objective function in the CTR prediction task is the binary cross entropy (BCE) loss~\cite{trench2013, chapelle2014simple, he2014practical}, which is defined as the cross entropy between the predicted click-through rate $\sigma(z_i)$ and the true label $y$. Mathematically,

\begin{align}
    \mathcal{L}_\text{BCE}=-\frac{1}{N}\sum_{i=1}^N [y_i \log(\sigma(z_i))+(1-y_i) \log(1-\sigma(z_i))],
\end{align}
where $N$ denotes the number of samples, $z_i$ represents the logit of $i$-th sample, and $y_i$ denotes the corresponding binary label, 1 for click and 0 for non-click. 

\subsection{Learning to Rank}
In scenarios such as contextual advertising, however, the pointwise approach often falls into sub-optimality. 
Firstly, the pointwise approach treats each document as an individual input object, disregarding the relative order between documents. 
Secondly, it fails to consider the query-level and position-based properties of evaluation measures for ranking~\cite{liu2009learning}. 
In contrast, learning-to-rank (LTR) methods can effectively address these issues and enhance ranking performance. 

Specifically, pairwise and listwise approaches are the two main branches of LTR.
Pairwise methods aim to ensure that the estimated value of positive samples is greater than that of negative samples for each pair of positive/negative samples. In this field, many effective works like Ranking SVM~\cite{joachims2002optimizing}, GBRank~\cite{zheng2007regression}, RankNet~\cite{ranknet} and PRM~\cite{pei2019personalized} have been proposed. Among them, RankNet stands out with its clean formulation that effectively captures the essence of pairwise comparisons, which is defined as:

\begin{align}
    \label{eq: ranknet}
    \mathcal{L}_\text{RankNet}=-\frac{1}{N^2}\sum\limits_{i=1}^N\sum\limits_{j=1}^N&[ y_{ij}\log(\sigma(z_i-z_j)) \\
    \nonumber
    +& (1-y_{ij})\log(1-\sigma(z_i-z_j))],
\end{align}
where $y_{ij} \in \{0,0.5,1\}$ corresponds to the conditions that $y_i<y_j, y_i=y_j,y_i>y_j$, respectively.
Following enhancements in the optimization process~\cite{burges2006learning} and the incorporation of hinge loss~\cite{tagami2013ctr} have led to further performance improvements. 

Listwise methods encourage positive samples to have higher rankings within the list of all samples. For example, ListNet~\cite{cao2007learning} defines its loss as:

\begin{align}
    \mathcal{L}_\text{ListNet}=-\frac{1}{N_+}\sum\limits_{i=1}^{N_+}\log \frac{\exp[z_{i}]}{\sum_{k=1}^N\exp[z_{k}]}.
    \label{listnet}
\end{align}

Some other listwise approaches have also achieved remarkable results, such as ListMLE~\cite{xia2008listwise}, BayesRank~\cite{kuo2009learning} and PiRank~\cite{swezey2021pirank}. 
% Recently, studies have revealed that such methods lack calibration capabilities, potentially leading to training instability. 
% Improved approaches like CalSoftmax~\cite{yan2022scale} have been proposed.

\subsection{Tailor Ranking Loss into CTR Prediction}

In the context of online advertising, recent studies~\cite{jrc,yan2022scale,bai2023regression} argued that it is not easy to achieve decent overall outcomes solely by relying on a single form of the objective function. 
Consequently, several works that combine classification loss and ranking loss have been proposed with the following objective paradigm:

\begin{align}
    \label{combine}
    \mathcal{L}=\alpha \mathcal{L}_\text{Clf}+(1-\alpha) \mathcal{L}_\text{Rank},
\end{align}  
where $\mathcal{L}_\text{Clf}$ and $\mathcal{L}_\text{Rank}$ are classification loss and ranking loss, respectively. 
Researchers have proposed a suite of approaches based on this optimization paradigm. 
Initial attempts were made with methods that combine Mean Squared Error with ranking loss~\cite{sculley2010combined} for the regression task. 
Subsequently, Combined-Pair~\cite{twitter}, being one of the pioneers, successfully integrated BCE loss with pairwise ranking loss in the field of industrial advertising, resulting in a consistent performance improvement. 

Inspired by Combined-Pair, several methods propose to combine BCE loss with other forms of ranking loss (\textit{e.g.}, hinge loss~\cite{yue2022learning}, triple-wise loss ~\cite{shan2018combined}), especially the listwise ranking loss. 
For example, Combined-List~\cite{yan2022scale} combines BCE loss with ListNet loss, RCR~\cite{bai2023regression} combines BCE loss with $\mathcal{L}_{rank}^{RCR}$, defined as:

\begin{align}
    \mathcal{L}_\text{Rank}^\text{RCR}=-\frac{1}{N_+}\sum\limits_{i=1}^{N_+}\log \frac{\sigma(z_{i})}{\sum_{k=1}^N\sigma(z_{k})},
\end{align}
where $\sigma(\cdot)$ represents sigmoid function. JRC~\cite{jrc} decouples the logit into click/non-click logits and is formalized as :

\begin{align}
    \label{jrc_eq}
    &\mathcal{L}_\text{Clf}^\text{JRC}=-\frac{1}{N}\sum\limits_{i=1}^N\log \frac{\exp [z_{iy_i}]}{\exp [z_{i0}]+\exp [z_{i1}]},\\
    & \mathcal{L}_\text{Rank}^\text{JRC}=-\frac{1}{N}\sum\limits_{i=1}^N\log \frac{\exp [z_{iy_i}]}{\sum_{k=1}^N\exp [z_{ky_i}]},
\end{align} 
where $z_{i1}$ and $z_{i0}$ are the click/non-click logits of the $i$-th sample, respectively. $z_{ky_i}$ is the click logit of $k$-th sample if $i$-th samples is positive. Otherwise, $z_{ky_i}$ signifies the non-click logit.
In Sec.~\ref{methametical_understanding} and Sec.~\ref{sec:empirical_analysis}, we will conduct detailed empirical experiments and analyses on these methods.
% \section{Ranking Loss from A Gradient Vanishing Perspective}
\section{Auxiliary Ranking Loss in CTR Prediction: A Gradient Vanishing Perspective}
\label{methametical_understanding}

The classification-ranking combination methods are widely used in real-world recommendation systems, \textit{e.g.}, Twitter~\cite{twitter}, Google~\cite{yan2022scale, bai2023regression}, and Alibaba~\cite{jrc}.
However, the effectiveness of such methods remains elusive, which drives us to explore the underlying mechanisms at play.
In the following, we study Combined-Pair loss~\cite{twitter} as a representative classification-ranking combination loss and choose DCN V2~\cite{wang2021dcn} as the backbone model.
The Combined-Pair loss is defined as a combination of a binary cross entropy (BCE) loss and a RankNet loss~\cite{ranknet}:

% \begin{align}
% \label{combined_pair}
%     \mathcal{L}^{CP} &= \alpha \mathcal{L}_\text{BCE} + (1-\alpha) \mathcal{L}_\text{rank}^{CP} \\
%     =&-[\alpha\cdot \frac{1}{N}\sum_{i\in N} [y_i \log(\sigma(z_i))+(1-y_i) \log(1-\sigma(z_i))] \nonumber\\
%      &+(1-\alpha)\cdot \frac{1}{N_+ N_-}\sum\limits_{i\in N_+} \sum\limits_{j\in N_-} \log(\sigma(z_i^{(+)}-z_j^{(-)}))],\nonumber
% \end{align} 

\begin{align}
    \label{combined_pair}
    &\mathcal{L}^\text{CP} = \alpha \mathcal{L}_\text{BCE} + (1-\alpha) \mathcal{L}_\text{RankNet}, \\
    % \mathcal{L}_\text{BCE} &= - \frac{1}{N}\sum_{i=1}^N [y_i \log(\sigma(z_i))+(1-y_i) \log(1-\sigma(z_i))], \\
    &\mathcal{L}_\text{RankNet} = -\frac{1}{N_+ N_-}\sum\limits_{i=1}^{N_+} \sum\limits_{j=1}^{N_-} \log(\sigma(z_i^{(+)}-z_j^{(-)})), 
\end{align} 
where $N_+$ and $N_-$ represent the numbers of positive and negative samples, respectively. RankNet within the Combined-Pair is a form of Eq.~\ref{eq: ranknet} without $y_i=y_j$.
In the following discussion, we name the model that optimizes the Combined-Pair loss as the Combined-Pair method, or Combined-Pair in short, and the model that optimizes only the BCE loss as the BCE method.

\subsection{Investigation of Classification Ability}

Existing research~\cite{twitter, jrc} posits that the BCE Loss mainly provides a decent estimate of click probability and involving a ranking loss in addition to the BCE loss improves its ranking performance.
% \zhutian{Existing work introduces pairwise approaches to enhance the ranking performance of impressions ordered by click probability~\cite{twitter, jrc}}.
% mainly attributes the success of classification-ranking combination loss to enhancing the ranking ability~\cite{twitter, jrc}.
However, we are curious about its impact on the main objective of CTR prediction: the \textit{classification capability}.
We train the BCE model and Combined-Pair on the public Criteo dataset~\cite{criteo-display-ad-challenge}.
Please note that the Criteo dataset's CTR is 25.6\%, which is relatively high due to downsampling on negative samples.
% still significantly differs from the online CTR of less than 2\%. 
A weight $\beta_\text{pos}$ for the positive samples was introduced during our model training and evaluation to simulate sparse positive rates in real-world scenarios.

We conducted monitoring of the \LossName{} for both the BCE method and the Combined-Pair on the validation set, as depicted by the dashed lines in Fig.~\ref{fig: finding_a}. 
Interestingly, we made an unexpected observation: the \LossName{} of the Combined-Pair method decreases more rapidly than that of the BCE method initially and maintains a consistently lower value thereafter. This finding can be summarized as follows:

\begin{tcolorbox}[colback=blue!2!white,leftrule=2.5mm,size=title]
    \emph{%
        Finding 1. Combined-Pair gets a lower \LossName{} than the BCE method on the validation set, indicating that it improves the classification ability rather than only the ranking ability.
    }
\end{tcolorbox}

To investigate whether the improvement in the classification loss is due to better generalization or easier model training when incorporating the ranking loss, we monitored the \LossName{} of both methods on the training set, as illustrated by the solid lines in Fig.~\ref{fig: finding_a}.
In addition to the previously mentioned surprising results, we made an even more astonishing observation: the \LossName{} of the Combined-Pair method on the training set also decreases more rapidly and remains consistently lower than that of the BCE method.
We conclude with the second finding:

\begin{tcolorbox}[colback=blue!2!white,leftrule=2.5mm,size=title]
    \emph{%
        Finding 2. Combined-Pair gets a lower \LossName{} than the BCE method on the training set, indicating that involving an auxiliary ranking loss helps the optimization of the \LossName{}.
    }
\end{tcolorbox}

Besides, we analyze the disparities between the loss landscapes of Combined-Pair and the BCE method. 
As shown in Fig.~\ref{fig: finding_c}, we observe that the loss landscape of the BCE method exhibits flatter than Combined-Pair.

\subsection{Gradients Analysis}
To gain further insights into the reasons behind the aforementioned observations, we conduct a detailed analysis of the gradients in both the BCE and Combined-Pair methods.
We begin by examining the gradients of the BCE method. 
According to the chain rule, the gradients of the parameters in each layer are proportional to the gradients of the logits. 
Hence, our initial focus is on studying the logit gradients.

\subsubsection{Gradients of BCE Loss for Negative Samples}
\label{section: bce_neg}

The gradient of BCE loss for negative sample $x_j$'s logit $z^{(-)}_j$ can be derived as:
% \begin{align}
%     \nabla_{z^{(-)}_j} \mathcal{L}_\text{BCE}=&\frac{1}{N}\cdot \frac{1}{1-\sigma(z^{(-)}_j)} \cdot 
%     \label{eq: neg_grad}
%      \nonumber
%      \sigma(z^{(-)}_j)(1-\sigma(z^{(-)}_j))\\
%     =& \frac{1}{N} \cdot \sigma(z^{(-)}_j) 
%     =\frac{1}{N} \cdot {\color{red}\hat{p}_{j}}.
% \end{align}

\begin{align}
    \nabla_{z^{(-)}_j} \mathcal{L}_\text{BCE}=& \frac{1}{1-\sigma(z^{(-)}_j)} \cdot 
    \label{eq: neg_grad}
     \nonumber
     \sigma(z^{(-)}_j)(1-\sigma(z^{(-)}_j))\\
    =& \sigma(z^{(-)}_j) 
    = {\color{red}\hat{p}_{j}}.
\end{align}

This equation demonstrates that the gradients of negative samples are proportional to its pCTR value, $\hat{p}_{j}$. 
The expected value of $\hat{p}_{j}$ produced by an unbiased CTR estimation model with BCE loss is close to the underlying global CTR, which equals approximately to the proportion of click samples (\textit{i.e.}, positive feedback) to the total samples. 
This is because the BCE loss function is \textit{scale calibrated}~\cite{yan2022scale} and its global minima are achieved at $\sigma(z)\rightarrow \mathrm{E}[y|x]$.

When the positive feedback is sparse (\textit{e.g.}, the CTR in our real-world advertising platform is usually less than $2\%$), $\hat{p_j}$ becomes a small value. 
According to Eq.~\ref{eq: neg_grad}, the gradients of negative samples are proportional to such \textit{small value} of $\hat{p_j}$ and tend to be relatively small. 
We refer to this as \textit{gradient vanishing of negative samples} under sparse positive feedback.
We conclude this finding as:

\begin{tcolorbox}[colback=blue!2!white,leftrule=2.5mm,size=title]
    \emph{%
        Finding 3. When positive feedback is sparse, the gradients of negative samples vanish since they are proportional to the estimated positive rates, which are small in an unbiased estimator.
    }
\end{tcolorbox}

\begin{figure}[!tp]
    \centering
    \subfigure {\includegraphics[width=.228\textwidth]{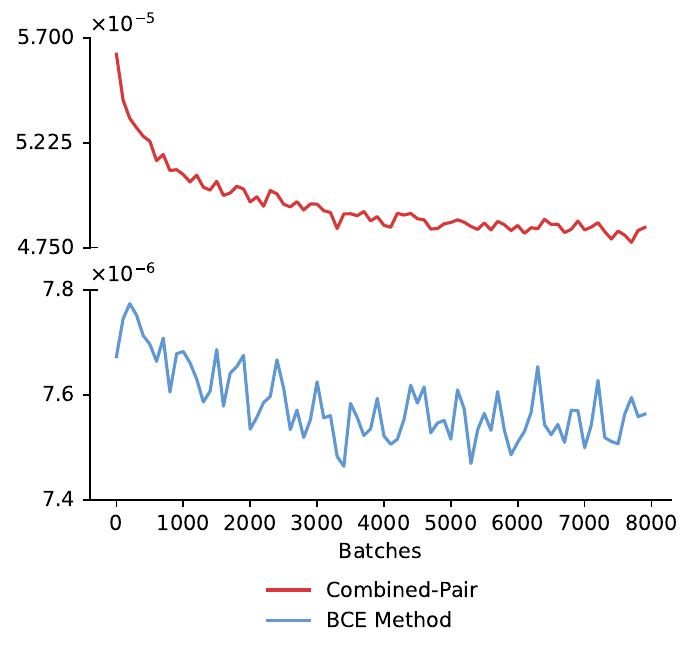}\label{fig: neg_grad_dual_0}}
    \subfigure {\includegraphics[width=.243\textwidth]{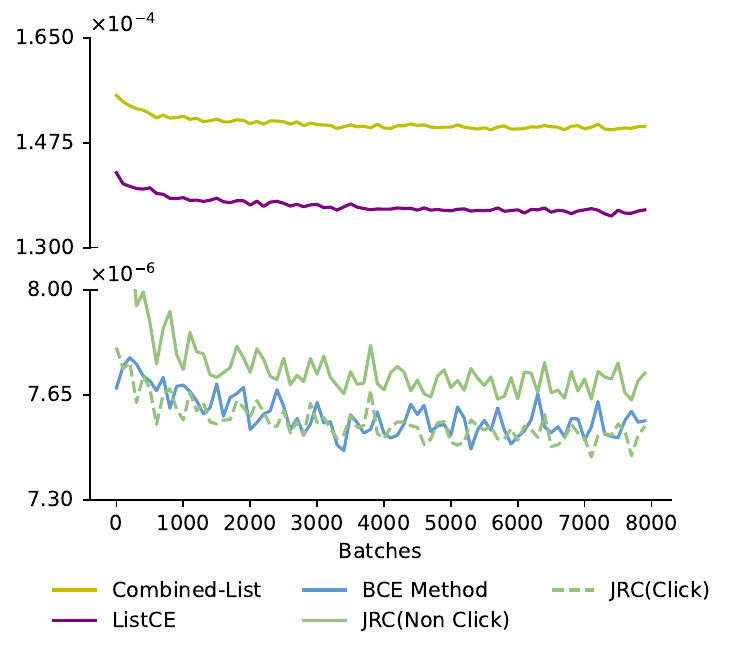}\label{fig: neg_grad_dual_1}}
    \caption{Gradient norm dynamics of negative samples logits in BCE method and BCE-pairwise ranking combination methods (left) and BCE-listwise ranking combination methods (right) on the Criteo dataset in the first training epoch. All methods set $\alpha=0.5$ in both plots.}
    \label{fig: dual_grad_comp}
\end{figure}

\subsubsection{Gradients of BCE Loss for Positive Samples}

We are curious whether positive samples also exhibit similar issues. As for a given positive sample $x_i$, the gradient of BCE for its logit $z^{(+)}_i$ can be derived as follows:

% \begin{align}
%     \nabla_{z^{(+)}_i} \mathcal{L}_\text{BCE}=&-\frac{1}{N}\cdot \frac{1}{\sigma(z^{(+)}_i)} \cdot 
%     \label{eq: pos_grad}
%      \nonumber
%      \sigma(z^{(+)}_i)(1-\sigma(z^{(+)}_i))\\
%     =& -\frac{1}{N}(1-\sigma(z^{(+)}_i))=-\frac{1}{N}\cdot ({\color{blue}1-\hat{p}_{i}}),
% \end{align}
\begin{align}
    \nabla_{z^{(+)}_i} \mathcal{L}_\text{BCE}=&-\frac{1}{\sigma(z^{(+)}_i)} \cdot 
    \label{eq: pos_grad}
     \nonumber
     \sigma(z^{(+)}_i)(1-\sigma(z^{(+)}_i))\\
    =& - (1-\sigma(z^{(+)}_i)) = - ({\color{blue}1-\hat{p}_{i}}).
\end{align}
According to Eq.~\ref{eq: pos_grad}, positive samples satisfy the $\nabla_{z^{(+)}_i} \mathcal{L}_\text{BCE}\propto 1-\hat{p_i}$, which is a relatively large value (close to $1$ when $\hat{p_i}$ is small) and therefore don't have gradient vanishing problems as negative samples do.

\subsubsection{Gradients of Combined-Pair for Negative Samples}
Combined-Pair contains two losses: BCE loss and RankNet loss. Here, We first discuss the gradients of the negative sample's logit in RankNet loss, which can be derived as:

% \begin{align}
%     \nabla_{z_j^{(-)}} \mathcal{L}_\text{Rank}^\text{CP} = \frac{1}{N_+ N_-} \sum\limits_{i\in N_+} \sigma(z_j^{(-)}-z_i^{(+)}).
% \end{align}
\begin{align}
    \nabla_{z_j^{(-)}} \mathcal{L}_\text{Rank}^\text{CP} = \frac{1}{N_+} \sum_{i=1}^{N_+} \sigma(z_j^{(-)}-z_i^{(+)}).
\end{align}

In both our online and offline advertising, when positive feedback is extremely sparse, it is observed that even the estimated values of positive samples tend to be much lower than 0.5. Consequently, the logit of positive samples $z_i^{(+)}$ is less than 0.  
This can result in greater gradients of negative samples in the RankNet Loss, compared to the BCE Loss, as follows:

% \begin{align}
%     \nabla_{z_j^{(-)}} \mathcal{L}_\text{Rank}^\text{CP} =& \frac{1}{N_+ N_-} \sum\limits_{i\in N_+} \sigma(z_j^{(-)}-z_i^{(+)})\\
%      \nonumber
%     >& \frac{1}{N_-}\cdot[\frac{1}{N_+}N_+\cdot \sigma(z_j^{(-)})] = \frac{1}{N_-}\sigma(z_j^{(-)}) \\
%      \nonumber
%     >&  \frac{1}{N}\sigma(z_j^{(-)})=\nabla_{z_j^{(-)}} \mathcal{L}_\text{BCE}.
% \end{align}

\begin{align}
    \nabla_{z_j^{(-)}} \mathcal{L}_\text{Rank}^\text{CP} =& \frac{1}{N_+} \sum_{i=1}^{N_+} \sigma(z_j^{(-)}-z_i^{(+)})\\
    \nonumber
    >& \frac{1}{N_+}\cdot N_+\cdot \sigma(z_j^{(-)}) \\
    \nonumber
    =& \sigma(z_j^{(-)}) = \nabla_{z_j^{(-)}} \mathcal{L}_\text{BCE}.
\end{align}

This indicates that in the sparse positive scenario and for the same negative sample logit, \textit{RankNet Loss may have larger gradients than BCE Loss}. 
Thus, the following inequation between the BCE method and the Combined-Pair method holds:

\begin{align}
    \nabla_{z_j^{(-)}} \mathcal{L}^\text{CP} =& \alpha \nabla_{z_j^{(-)}} \mathcal{L}_\text{BCE}+(1-\alpha)\nabla_{z_j^{(-)}} \mathcal{L}_\text{Rank}^\text{CP}\\
     \nonumber
    >& \alpha \nabla_{z_j^{(-)}} \mathcal{L}_\text{BCE}+(1-\alpha)\nabla_{z_j^{(-)}} \mathcal{L}_\text{BCE} \\
    \nonumber
    =& \nabla_{z_j^{(-)}} \mathcal{L}_\text{BCE}.
\end{align}

We conclude with the following finding:
\begin{tcolorbox}[colback=blue!2!white,leftrule=2.5mm,size=title]
    \emph{%
        Finding 4. When positive feedback is sparse, Combined-Pair has \textit{larger gradients} for negative samples than the BCE method. 
    }
\end{tcolorbox}

\subsection{Empirical Analysis of Gradient Vanishing}
To empirically validate the analysis, we examine the dynamics of gradient norms for negative sample logits in the first epoch of training. 
To simulate the low proportion of positive samples often encountered in real-world scenarios, we adjust the positive sample rates, denoted as $\beta_\text{pos}$, to achieve an equivalent Click-Through Rate (CTR) of 3.3\% for the dataset. 
Further details regarding this adjustment will be discussed in Sec.~\ref{subsec:setting}.

As depicted in Fig.~\ref{fig: dual_grad_comp}, we observe that the BCE method (the blue line) exhibits negligible gradient norms for negative samples. In contrast, the Combined-Pair method (the red line) demonstrates significantly larger gradient norms for negative samples. 
This empirical observation aligns with our previous analysis of the gradients.

\begin{figure}[!htbp]
    \centering
    \subfigure {\includegraphics[width=.23\textwidth]{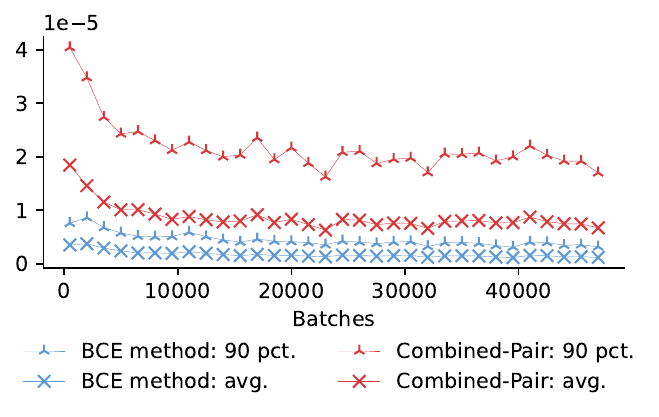}\label{fig: layers_grad_dnn}}
    \subfigure {\includegraphics[width=.23\textwidth]{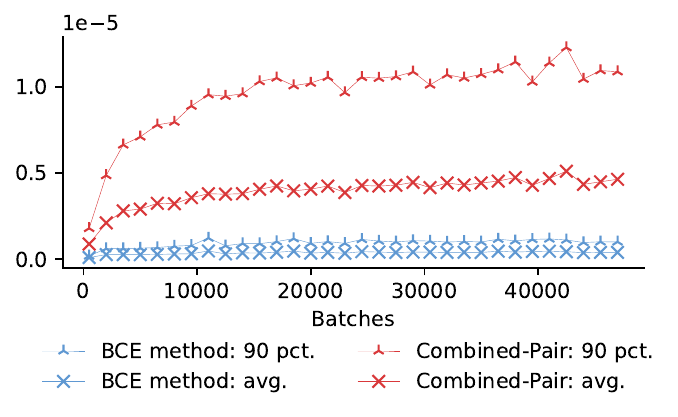}\label{fig: layers_grad_dnn}}
    \caption{Gradient norm dynamics of DNN (left) and CrossNet (right) in DCN V2 through the training process. pct and avg. are shorts for percentile and average.}
    \label{fig: layers_grad}
\end{figure}

We further investigate the optimization procedure of the trainable parameters in the entire model architecture. 
Specifically, we compare the Combined-Pair method with the BCE method by examining the gradient norms of the trainable parameters in the bottom layers of the Deep Neural Network (DNN) and CrossNet in DCN V2.
To provide a comprehensive understanding of the optimization dynamics, we report the dynamics of the 90th percentile and average values of the gradient norms during the training process. 
These results are depicted in Fig.~\ref{fig: layers_grad}.
Remarkably, we observe that the Combined-Pair method consistently achieves higher values on both metrics compared to the BCE method. 
This difference persists throughout the entire training process. 
This finding further validates that the Combined-Pair method effectively alleviates the issue of gradient vanishing in the learnable parameters.

\section{Experiments}
\label{sec:empirical_analysis}

In this section, we conduct empirical experiments to answer the following research questions (RQs): 
If the sparse positive rate is the main cause of the gradient vanishing of negative samples, how would models perform under various sparsity of positives? 
% Does the Combined-Pair get a larger performance lift on datasets with sparser positives (RQ1)? 
What's the trade-off between classification loss and ranking loss (RQ2)?
How do the other methods with classification-ranking combination losses perform (RQ3)?
Last, can our perspective be extended to methods beyond the classification-ranking combination loss (RQ4)? 
% Can other methods also alleviate the gradient vanishing issue and hence improve the classification performance (RQ4)?

% they also alleviate the gradient vanishing and hence improve performance (RQ4)?
% First, we already show that the Combined-Pair outperforms the BCE method since it mitigates the gradient vanishing issue when the positive feedback is sparse. 
% So, if we create artificial datasets with various positive sparsity rates, how would these two methods perform (RQ1)?

\subsection{Dataset and Experimental Setting}
\label{subsec:setting}
We conducted experiments on the public Criteo dataset\footnote{https://github.com/reczoo/Datasets/tree/main/Criteo/Criteo\_x1}~\cite{criteo-display-ad-challenge}, a widely used advertising recommendation dataset. 
It consists of 13 numerical features and 26 categorical features. Specifically, we utilized the criteo\_x1 version, where the training, validation, and testing data are divided in a 7:2:1 ratio.
We employed DCN V2~\cite{wang2021dcn} as the backbone model and utilized the FuxiCTR~\cite{zhu2021open} implementation\footnote{https://github.com/reczoo/FuxiCTR/tree/main/model\_zoo/DCNv2} with the same settings as BARS~\cite{zhu2022bars}\footnote{https://github.com/reczoo/BARS/tree/main/ranking/ctr/DCNv2/DCNv2\_criteo\_x1}.
We evaluated the performance regarding two metrics: \LossName{} (\textit{i.e.}, binary-cross entropy loss) to measure the classification ability and AUC to measure the ranking ability.
% AUC of ROC to measure the ranking ability.

Specifically, we create artificial datasets based on the Criteo dataset and control the sparsity degree of positives by assigning a weight $0< \beta_\text{pos} \leq 1$ for all its positive samples.
This weight is used to down-weight positive samples in the training loss:

\begin{equation}
    \mathcal{L}_{BCE}=-\frac{1}{N}\sum_{i=1}^N [\beta_\text{pos} \cdot y_i \log(\sigma(z_i))+(1-y_i) \log(1-\sigma(z_i))].
\end{equation}

For example, by setting $\beta_\text{pos} = 0.1$ for all positive samples, we generate a new dataset with sparsity degree of positive as $\frac{25.6\%\times 0.1}{25.6\%\times 0.1+1-25.6\%}=3.3\%$.

\subsection{RQ1: Performance Evaluation with various Positive Sparsity Rates}
\label{subsec:various_pos_sparsity_rates}

To validate our hypothesis, we generated multiple artificial datasets with varying degrees of sparsity in positive feedback by adjusting the values of $\beta_\text{pos}$. 
We then evaluated the performance of both the BCE method and the Combined-Pair method on these datasets.
If our hypothesis holds true, datasets with sparser positive feedback should be more susceptible to the issue of gradient vanishing. 
Consequently, we would expect the BCE method to exhibit poorer performance on these datasets, while the Combined-Pair method should demonstrate significantly better performance, resulting in a larger performance gap compared to the BCE method.

\begin{figure}[!tp]
    \centering
    \subfigure[AUC] {\includegraphics[width=.24\textwidth]{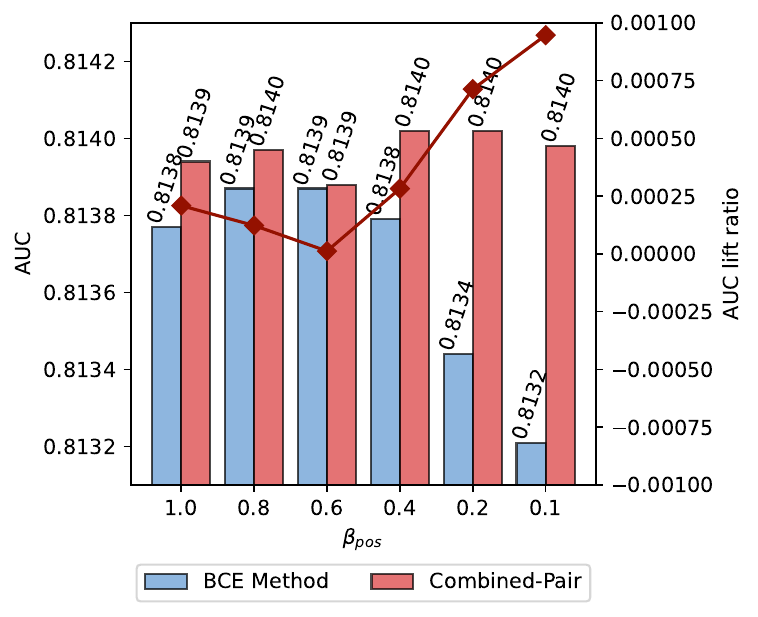}\label{fig: layers_grad_dnn}}
    \subfigure[\LossName{}] {\includegraphics[width=.227\textwidth]{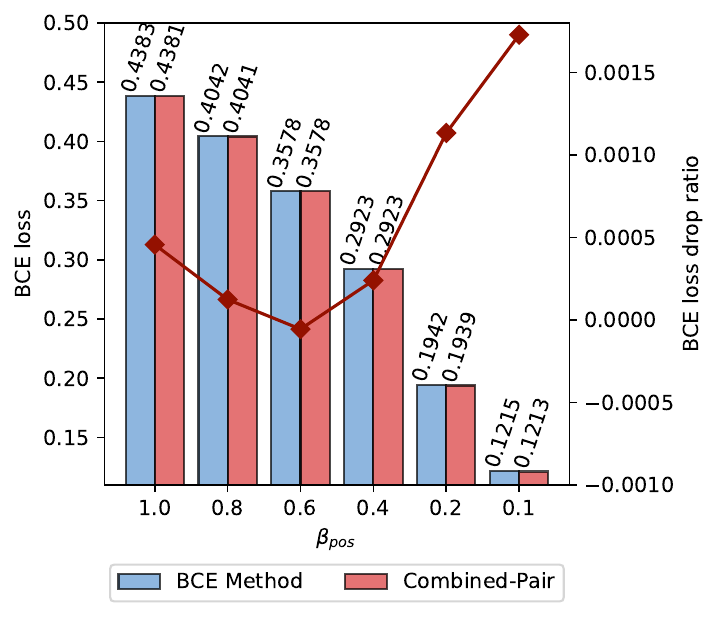}\label{fig: layers_grad_dnn}}
    \caption{Performance evaluation of Combined-Pair and the BCE method under varying positive sparsity rates. Here $\beta_\text{pos}$ denotes the weights of positive samples.}
    \label{fig:auc_\LossName{}}
\end{figure}

In particular, we created artificial datasets with $\beta_\text{pos}$ equals to $0.8$, $0.6$, $0.4$, $0.2$, $0.1$, respectively.
A smaller $\beta_\text{pos}$ indicates sparser positive feedback.
As shown in Fig.~\ref{fig:auc_\LossName{}}, we observe that the Combined-Pair always gets better AUC and BCE loss at all $\beta_\text{pos}$.
Especially, from $\beta_\text{pos}=0.6$ to $\beta_\text{pos}=0.1$, \emph{the sparser positive rates (\textit{i.e.}, smaller $\beta_\text{pos}$), the larger AUC lift (from 0.020\% to 0.095\%) and \LossName{} drop (from 0.045\% to 0.168\%) between the Combined-Pair and the BCE method}.
This validates our hypothesis that Combined-Pair achieves a larger performance lift than the BCE method when sparser positives reach a sparsity threshold (here $\beta_\text{pos}=0.6$).

\subsection{RQ2: Trade-off between Classification and Ranking Loss}
\label{subsec:class_vs_ranking}

% \cmt{re-write}
% We aim to investigate the trade-off between classification and ranking loss within the Combined-Pair loss.
% Specifically, we tune the loss weight $\alpha$ in Conbined-Pair from $1.0$ to $0.1$ and evaluate both negative \LossName{} and AUC, as shown in Fig.~\ref{fig: res_on_beta}. 
% We have the following observation.

% First, starting with $\alpha=1.0$, which equals to the BCE method (the red square), when we decrease $\alpha$, \textit{i.e.}, reduce the weight of the BCE loss while increasing the weight for the ranking loss, both negative \LossName{} and AUC are first improved simultaneously, represented by the orange arrow.
% For example, with $\beta_\text{pos}=0.1$, decreasing $\alpha$ from $1.0$ to $0.7$ leads to an improvement of negative \LossName{} from $0.1215$ to $0.1213$, and AUC from $0.8132$ to $0.8139$.
% This indicates that the classification and ranking ability can be both improved monotonically along with decreasing $\alpha$ to a threshold.

% Second, both metrics deteriorate when the $\alpha$ is further increased.
% That is to say, when the ranking loss becomes more dominant in the combination loss after a threshold, both classification and ranking ability deteriorate monotonically when the weight of the ranking loss becomes larger.
% However, when the positive is very sparse, \textit{e.g.}, $\beta_\text{pos}=0.1$, even with a very large weight for the ranking loss, \textit{e.g.}, $1-\alpha=0.9$, the model's performance is still better than the BCE method.

% ---

\begin{figure}[!tp]
    \centering
    \includegraphics[width=0.45\textwidth]{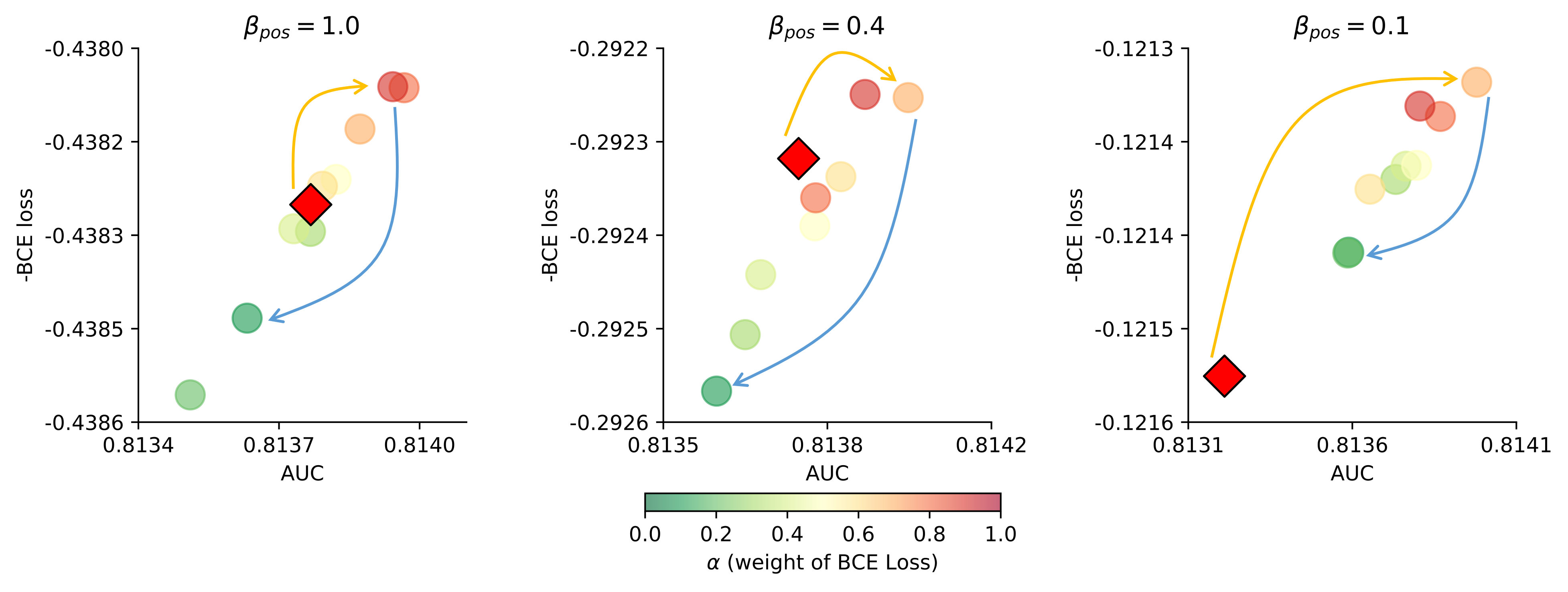}
    \caption{AUC and negative \LossName{} of Combined-Pair and BCE method in Criteo test set. $\alpha$ and $1-\alpha$ are the weights of BCE loss and RankNet loss within Combined-Pair, respectively. The diamond in red represents the BCE method, \textit{i.e.}, $\alpha=1$. We shows results with $\alpha$ ranging between $[0.1,1]$}
    \label{fig: res_on_beta}
\end{figure}

Our aim to examine the trade-off between the classification and ranking loss components within the Combined-Pair loss. To achieve this, we vary the loss weight parameter, denoted as $\alpha$, in the Combined-Pair loss from $1.0$ to $0.1$. We evaluate the negative \LossName{} and the Area Under the Curve (AUC) as performance metrics, as depicted in Fig.~\ref{fig: res_on_beta}. Based on our observations, we note the following:

First, starting with $\alpha=1.0$, which corresponds to the Binary Cross-Entropy (BCE) method (represented by the red diamond), we observe that decreasing $\alpha$, \textit{i.e.}, reducing the weight of the BCE loss while increasing the weight for the ranking loss, leads to simultaneous improvements in both the negative \LossName{} and AUC. This trend is represented by the orange arrow in the figure. For instance, with $\beta_\text{pos}=0.1$, reducing $\alpha$ from 1.0 to 0.7 results in a decrease in negative \LossName{} from 0.1215 to 0.1213, and an increase in AUC from 0.8132 to 0.8139. This suggests that \emph{the classification and ranking abilities can be improved monotonically by decreasing $\alpha$ up to a certain threshold}.

Second, we observe that both metrics deteriorate when $\alpha$ is further increased. In other words, as the ranking loss becomes more dominant in the combination loss beyond a certain threshold, both the classification and ranking abilities deteriorate monotonically (as shown by the blue arrow).

However, when the positive feedback is very sparse (\textit{e.g.}, $\beta_\text{pos}=0.1$), even with a very large weight for the ranking loss (\textit{e.g.}, $1-\alpha=0.9$), the model's performance remains superior to the BCE method.

\subsection{RQ3: Evaluation of Different Ranking Losses}
\label{subsec:different_ranking_loss}
In this section, we expand our analysis to include other classification-ranking combination methods beyond Combined-Pair. 
We aim to examine whether these methods possess properties similar to the Combined-Pair, thereby enhancing the applicability of our theory to a broader range of methods.

While Combined-Pair integrates BCE loss with pairwise ranking loss, other methods combine BCE loss with listwise ranking loss. 
For example, Combined-List~\cite{yan2022scale} employs BCE loss in conjunction with the original ListNet loss~\cite{cao2007learning}. 
RCR~\cite{bai2023regression} combines BCE loss with ListCE loss, a variant of ListNet loss designed to align its minima with that of BCE loss.
JRC~\cite{jrc} decoupled the logit into click and non-click logits and proposed a corresponding combination method with listwise-like loss. 

These combination methods also incorporate ranking loss with BCE loss. 
So, can they also alleviate gradient vanishing regarding negative samples? 
We monitor the gradient norm dynamics of negative samples for these methods on the Criteo dataset in the first training epoch.
Through the analysis shown in Fig.~\ref{fig: dual_grad_comp}, we found that their gradient norms are relatively improved compared to the BCE method to varying degrees across different forms, indicating that gradient vanishing is also alleviated in these methods.
% alleviation also occurs in combination methods other than Combined-Pair. 
Among them, JRC decouples the logit into click and non-click logit, and the gradient vanishing is mainly mitigated on the non-click logit.

% since the JRC decouples the logit into click and non-click logits, their gradient magnitudes differ. 
% Thus, JRC

% has different gradient norms between click/non-click logits.
% Empirically, the gradient of non-click logit is much larger, and therefore, the role in alleviating gradient vanishing mainly comes from non-click logit.

\begin{table}[!h]
    \caption{The performance of combining different ranking losses under sparse positive feedback situations. The $\uparrow$ and $\downarrow$ represent increasing in AUC and decreasing in \LossName{} compared to the BCE method, respectively.}
    \label{tab: bce_jrc_twitter}
    \resizebox{0.95\linewidth}{!}{
    \begin{tabular}{@{}l|c|c|ccc@{}}
    \toprule
    \multicolumn{1}{c|}{\multirow{2}{*}{\textbf{Metric}}} & \multirow{2}{*}{\textbf{BCE}} & \textbf{BCE+Pairwise} & \multicolumn{3}{c}{\textbf{BCE+Listwise}} \\ \cmidrule(l){3-6} 
    \multicolumn{1}{c|}{}                                 &                               & Combined-Pair         & JRC       & Combined-List      & RCR      \\ \midrule
    AUC$\uparrow$  &0.81321& \textbf{0.81398}$^\uparrow$   & 0.81355$^\uparrow$  & 0.81351$^\uparrow$        & 0.81349$^\uparrow$ \\
    \LossName{}$\downarrow$ & 0.12152 & \textbf{0.12131}$^\downarrow$  & 0.12146$^\downarrow$  & 0.12152       & 0.12141$^\downarrow$  \\\bottomrule
    \end{tabular}
    }
\end{table}

% Furthermore, does their alleviating gradient vanishing also lead to performance improvement? 
We then compare the performance of JRC, Combined-List, and RCR methods on the Criteo dataset with $\beta_\text{pos}=0.1$.
We found they all show improved ranking and classification performance, as shown in Tab.~\ref{tab: bce_jrc_twitter}. 
Overall, it can be concluded that these combination methods can also achieve performance improvement by introducing ranking loss to alleviate gradient vanishing of negative samples.

\begin{figure}[!tp]
    \centering
    \subfigure[$\beta_\text{pos}=1.0$] {\includegraphics[width=.155\textwidth]{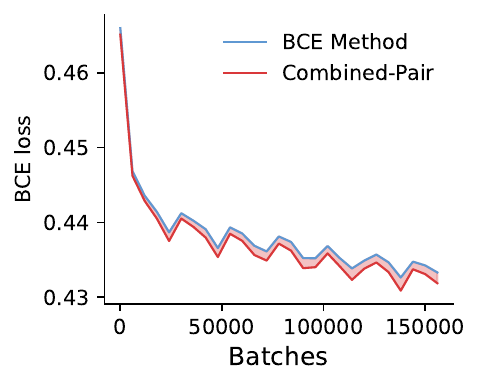}\label{fig: loss_cmp_along_beta1}}
    \subfigure[$\beta_\text{pos}=0.4$] {\includegraphics[width=.155\textwidth]{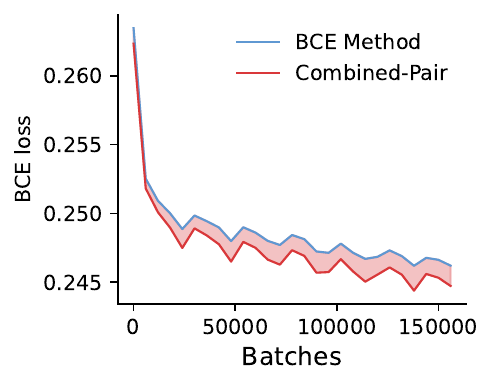}\label{fig: loss_cmp_along_beta0.4}}
    \subfigure[$\beta_\text{pos}=0.1$] {\includegraphics[width=.155\textwidth]{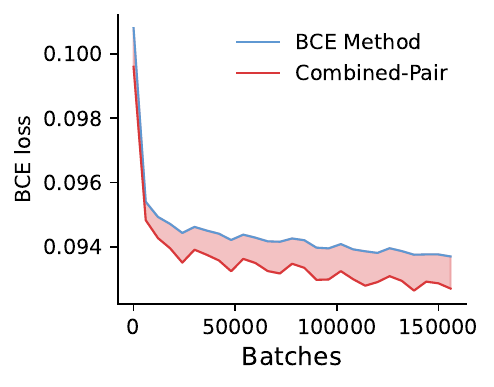}\label{loss_cmp_along_beta0.1}}
    \caption{Training error (\textit{i.e.}, \LossName{}) of the BCE method and Combined-Pair with various $\beta_\text{pos}$ on Criteo.}
    \label{fig: loss_cmp_along_beta}
\end{figure}

\subsection{RQ4: Beyond Ranking Loss}
\label{subsec:beyongd_ranking_loss}

We're curious whether approaches beyond ranking loss can also alleviate the gradient vanishing of negative samples and hence improve classification performance.
We first study the following two methods: Focal Loss~\cite{lin2017focal} and Negative Sampling~\cite{mikolov2013distributed}.
Then, we designed a novel approach called Combined-Contrastive to validate our perspective further.

\subsubsection{Focal Loss}

It assigns higher weights to poorly classified samples~\cite{lin2017focal}, \textit{i.e.}, negative samples suffering from gradient vanishing in our scenario.
Specifically, Focal Loss introduces a weight with hyper-parameter $\gamma$ to control the weight of samples:

\begin{align}
    \mathcal{L}_\text{Focal}=-\frac{1}{N}\sum_{i=1}^N [y_i (1-\hat{p}_i)^\gamma \log(\hat{p}_i)+(1-y_i)\hat{p}_i^\gamma \log(1-\hat{p}_i))],
\end{align}
where $\gamma$ controls the relative weight.
For those negative samples that may suffer from gradient vanishing when only using the BCE loss, their prediction score $\hat{p}_i$ should be wrongly high, making them have a higher weights $\hat{p}_i^{\gamma}$ than those negatives that are well-classified, \textit{i.e.}, with a low score.
The larger $\gamma$, the higher the relative weights to those poorly-classified samples.

\begin{figure}[!tbp]
    \centering
    % \subfigure[Focal Loss] {\includegraphics[width=.26\textwidth]{figure/focal.pdf}\label{fig: focal_grad}}
    % \subfigure[Negative Sampling] {\centering\includegraphics[width=.211\textwidth]
    \subfigure {\includegraphics[width=.241\textwidth]{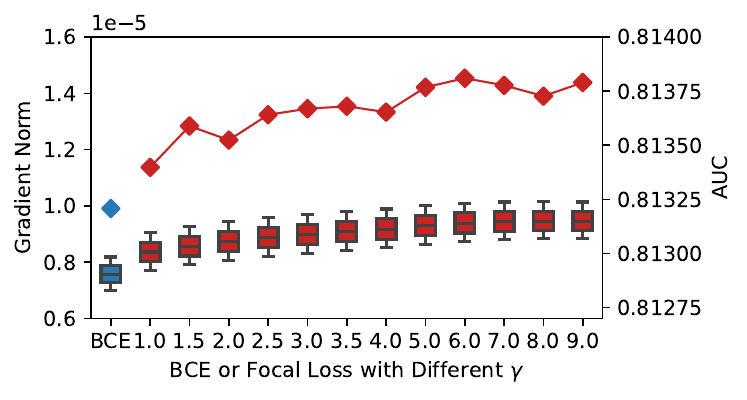}\label{fig: focal_grad}}
    \subfigure {\centering\includegraphics[width=.228\textwidth]{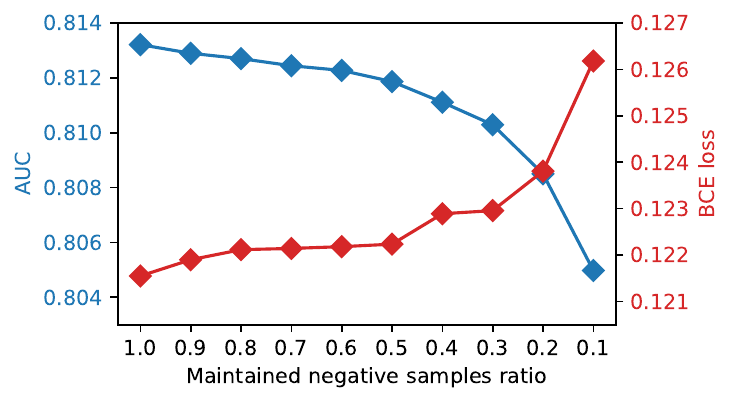}\label{fig: neg}}
    \caption{Evaluation of Focal loss and negative sampling.
    Left: Gradient norm (boxplot) and AUC (solid line) along with different $\gamma$ for Focal Loss.
    Right: AUC (blue) and \LossName{} (red) after isotonic regression with increasingly aggressive Negative Sampling.}
    \label{fig: beyond_rk}
\end{figure}

We then conducted a set of experiments by comparing gradients and performance for Focal Loss with different $\gamma$. 
As shown in Fig.~\ref{fig: beyond_rk} (left), similar to Combined-Pair, Focal Loss also gets higher gradients on negative samples than the BCE loss.
The larger the $\gamma$, the more weights on the poorly classified samples, and the larger the performance lift than the BCE method. 
These results validate that Focal Loss can also mitigate the gradient vanishing of negative samples and, hence, improve classification performance.
% further validate our perspective that the gradient vanishing of negative samples is the key challenge when positives are sparse; therefore, resolving it with either ranking loss or Focal Loss leads to performance lift.

Please note that the original $\hat{p_i}^\gamma$ is always less than $1.0$, and the gradients of negative samples in Focal loss are constantly smaller than BCE loss. 
For a fair comparison, we introduce a slight modification to the original formulation by replacing  $\hat{p_i}^\gamma$ with $[\hat{p_i}^\gamma + (1-\sum_{k\in N_-}\hat{p_k}^\gamma/N_-)]$. 
This adjustment normalizes the average weights of the negative samples to $1.0$, aligning it with the BCE loss. 
Formally, the modified version is defined as:

\begin{align}
\mathcal{L}_\text{Focal'}=-\frac{1}{N}&\sum\limits_{i=1}^N [y_i (1-\hat{p}_i)^\gamma \log(\hat{p}_i)\\
\nonumber
+&[\hat{p_i}^\gamma + (1-\sum_{k=1}^{N_-}\hat{p_k}^\gamma/N_-)](1-y_i) \log(1-\hat{p}_i))],
\end{align}

\subsubsection{Negative Sampling}

Another trial is to reduce the proportion of negative samples through negative sampling~\cite{mikolov2013distributed}, thereby increasing the estimated CTR.
This may consequently increase the gradient of negative samples because their gradients are proportional to the estimated CTR (Eq.~\ref{eq: neg_grad}).
% However, optimizing quality is not only related to gradient issues but also to the quality of negative samples. 
% Some studies~\cite{chen2020simple} have shown that negative samples of appropriate scale and learning difficulty contribute to learning representations of uniformity. 
However, downsampling negative samples may lead to information degradation, especially in datasets like Criteo, which has already undergone negative sampling.
We analyze the AUC and \LossName{} of negative sampling, and for a fair comparison, we report the calibrated results by isotonic regression~\cite{chakravarti1989isotonic,leeuw2009isotone}.
We observe that as negative sampling becomes more aggressive, both the model's ranking and classification capabilities deteriorate in Fig.~\ref{fig: beyond_rk} (right), which indicates that negative sampling fails to improve performance.

\subsubsection{Combined-Contrastive: A New Method}
% Upon the discussion on previous sections, our perspective is validated on various existing methods. 
% Furthermore, we wonder whether our perspective can also provide guidance for a novel method proposal.
% Inspired by ~\cite{khosla2020supervised}, we designed a new method, Combined-Contrastive Loss, which combines the BCE loss with the Contrastive Loss.
% Detailedly, the Contrastive Loss aims to push the embeddings belong to the same class closely, while separate embeddings from different classes distinctly.
% The Combined-Contrastive Loss is defined as:

% Based on the discussions presented in the previous sections, our perspective is substantiated by the validation against various existing methods. Moreover, we aim to investigate the applicability of our perspective in guiding the proposal of a novel method.

Besides validating our perspective on existing methods, such as Combined-Pair and Focal Loss, we'd like to derive new methods based on our perspective.
We speculate that introducing an auxiliary loss that considers the label information, \textit{i.e.}, an \emph{auxiliary supervised loss}, may provide larger gradients than the Binary Cross-Entropy (BCE) loss itself and hence mitigate the gradient vanishing issue.
To this end, inspired by the supervised contrastive learning~\cite{khosla2020supervised}, we have devised a novel approach termed Combined-Contrastive Loss, which integrates the BCE loss with Contrastive Loss. 
Specifically, the Contrastive Loss is employed to encourage embeddings belonging to the same class to be closely grouped while ensuring distinct separation between embeddings from different classes. Formally,

\begin{align}
    &\mathcal{L}^\text{CC} = \alpha \mathcal{L}_\text{BCE}+ (1-\alpha) \mathcal{L}_\text{Contr},\\
    &\mathcal{L}_\text{Contr}=\frac{1}{|N|}
    \sum_{i=1}^N \frac{-1}{|P(i)|} \sum_{p \in P(i)} \log \frac{\exp (\mathbf{z}_i\mathbf{z}_p/ \tau)}{\sum\limits_{a \in A(i)} \exp (\mathbf{z}_i\mathbf{z}_a/ \tau)},
\end{align}
where $\mathcal{L}_\text{CC}$ and $\mathcal{L}_\text{Contr}$ represent the Combined-Contrastive Loss and the Contrastive Loss, respectively.
$N$ denotes the number of samples in the batch, $A(i)$ denotes the whole samples set except $i$-th sample itself, $P(i)$ denotes a sample subset that contains all samples in $A(i)$ that are with the same label as $i$-th sample, $\mathbf{z_i}$ denotes the embedding of the $i$-th sample. We set $\alpha=0.9$ and $\tau=0.4$.

\begin{table}[!ht]
    \caption{Performance and gradient norm of negative samples of BCE method and Combined-Contrastive. The training stage's average values on the first epoch are reported.}
    \label{tab: cc}
    \resizebox{0.99\linewidth}{!}{
    \begin{tabular}{@{}l|l|cc@{}}
    \toprule
    \multicolumn{1}{c|}{\textbf{Stage}} & \multicolumn{1}{c|}{\textbf{Metrics}} & \textbf{BCE Method} & \textbf{Combined-Contrastive} \\ \midrule
    \multirow{2}{*}{Training}           & Gradient Norm                         & $4.9\times 10^{-6}$ & $\mathbf{7.5 \times 10^{-6}}$ \\
                                        & \LossName{} $\downarrow$                             & 0.09667             & \textbf{0.09428}$^\downarrow$              \\ \midrule
    \multirow{2}{*}{Testing}            & AUC$\uparrow$                                   & 0.81321             & \textbf{0.81340}$^\uparrow$              \\
                                        & \LossName{}$\downarrow$                               & 0.12152             & \textbf{0.12147}$^\downarrow$              \\ \bottomrule
    \end{tabular}
    }
\end{table}

We conduct experiments on the same artificial Criteo dataset with $\beta_\text{pos}=0.1$ as mentioned in Sec.~\ref{subsec:setting}.
As shown in Tab.~\ref{tab: cc}, Combined-Contrastive gets a higher AUC and smaller \LossName{} in the testing set, indicating better classification and ranking ability.
In addition, similar to the Combined-Pair, it also gets lower training \LossName{} and larger gradients than the BCE method.
This verifies that by introducing the auxiliary contrastive loss, Combined-Contrastive can also mitigate the gradient vanishing issue and hence improve the classification ability.

% Unlike the combination loss of BCE Loss and ranking loss, the Combined-Contrastive does not introduce any ranking information.
% We wonder whether Combined-Contrastive has only better generalization ability or a better optimization process.
% We then analyze the training error, \textit{i.e., }\LossName{} in the training stage, and find a smaller \LossName{} than the BCE method. 
% This indicates that the Combined-Contrastive shows a better optimization process.

% Furthermore, we compare the gradient norm between the two methods on negative samples' embeddings.
% , since the contrastive loss acts on embedding-level. 
% We observe that Combined-Contrastive shows a greater gradient norm than the BCE method.
% \zhutian{Such a greater gradient can be attributed to the fact that even a pair of negative samples can still offer a gradient by contrastive loss. 
% Consequently, sufficient gradients can be provided even in sparse positive feedback. In contrast, BCE Loss is susceptible to sparsity, which vanishes gradients for negative samples.}

% These further support that Combined-Constrastive can improve optimization by alleviating gradient vanishing in negative samples, thus achieving better classification and ranking performance.
% Hence, the 
% Combined-Contrastive shows a higher gradient norm than BCE method.
% It supports that introducing Contrastive Loss can alleviate the issue of gradient vanishing for negative samples in BCE Loss.

\section{Stability and Compatibility}
\label{sec:stability_and_compatibility}

Recent studies\cite{bai2023regression,yan2022scale} have raised concerns regarding tailoring ranking loss into CTR prediction. 
Some studies~\cite{yan2022scale} are concerned that ranking loss introduces \textit{score drifting}, leading to \textit{optimization instability}, while others~\cite{bai2023regression} express concerns about \textit{compatibility} issues between ranking loss and BCE loss.
Both issues can also lead to negative effects on optimization. 
Hence, we aim to investigate whether classification-ranking loss, especially Combined-Pair, also has these two issues or not.

\subsection{Stablility}
\label{stability}
Stability~\cite{yan2022scale} refers to the phenomenon that the model scores may keep drifting during model training.
Intuitively, singly employing ranking loss causes the model to focus solely on the relative orders between samples, neglecting the absolute prediction scores. 
This leads to \textit{score drifting}~\cite{yan2022scale}, resulting in non-scale calibrated outcomes and exacerbating bias distribution.
Hence, we wonder whether combining ranking loss with BCE loss in Combined-Pair may worsen the bias compared to using only BCE loss (\textit{i.e.}, the BCE method). 
It is worth mentioning that we did not employ any post-processing calibration techniques here. Otherwise, the bias would not reflect its training stability.

\begin{figure}[!ht]
	\centering
	\subfigure[Offline Bias] {\includegraphics[width=.22\textwidth]{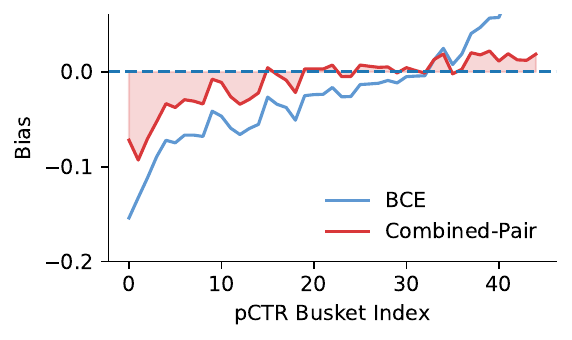}}
	\subfigure[Online Bias] {\includegraphics[width=.23\textwidth]{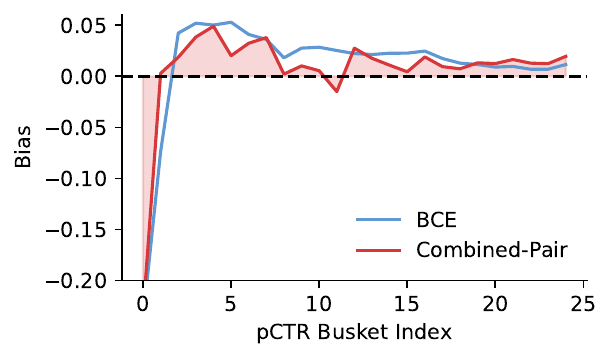}}
	\caption{The bias distribution over different pCTR buskets for both online and offline experiments.}
	\label{fig: bias}
\end{figure}

We bucketize the samples with equal frequency of positive samples and plot the bias of corresponding samples within each bucket. 
As shown in Fig.~\ref{fig: bias}, in both online and offline advertising, the BCE method severely underestimates the click-through rate (CTR) in the lower buckets, while Combined-Pair has a much smaller bias. 
In the higher buckets, the BCE method overestimates the CTR, and Combined-Pair also has a small bias in those buckets.
The reason is probably that Combined-Pair mitigates the gradient vanishing issue, thus facilitating model optimization and leading to lower bias.
In summary, Combined-Pair demonstrates \textit{superior calibration ability} compared to the BCE method, avoiding the issue of score drifting and ensuring stability.

\subsection{Compatibility}
Recent research~\cite{bai2023regression} argued that the pointwise and pairwise loss may not be compatible since they have different global minima.
Thus, we are curious about the \textit{compatibility} of the two losses within Combined-Pair: the BCE loss vs. the RankNet loss.
We refrain from comparing their global minima~\cite{bai2023regression}, as differences in global minima among losses do not inherently indicate incompatibility. 
For instance, while \textit{l1} or \textit{l2} regularization losses do not share the same global minimum as most optimization objectives, they are still effectively utilized in tandem to mitigate overfitting.

\begin{table}[!ht]
\caption{Gradient of positive and negative sample logit.}
\label{tab: compatibility}
\begin{tabular}{l@{}|cc|c@{}}
\toprule
\multicolumn{1}{l|}{\textbf{Samples}} & \textbf{BCE Loss}                   & \textbf{RankNet Loss}                                          & \multicolumn{1}{c}{\textbf{Direction}} \\ \midrule
Negative                & $\frac{\hat{p}_{j}}{N} > 0 $                     & $\frac{\sum_{i=1}^{N_+} \sigma(z_j^{(-)}-z_i^{(+)})}{N_+ N_-} >0$ & Same                 \\
Positive                & $-\frac{1-\hat{p}_{i}}{N} <0$ & $-\frac{\sum_{j=1}^{N_-}[1-\sigma(z_i^{(+)}-z_j^{(-)})]}{N_+ N_-} <0$                    & Same                    \\ \bottomrule
\end{tabular}
\end{table}

Instead, we investigate the compatibility of these losses by analyzing their gradients, examining whether there is a conflict in gradients between them.
As shown in Tab.~\ref{tab: compatibility}, the BCE loss and RankNet loss gradients are always aligned in the same sign, leading to the same optimization directions by two objectives. Therefore, the optimization directions of the combined-pair loss remain compatible without conflict.

\section{Online Deployment}
\label{online}
% We empirically evaluate Combined-Pair on the CTR prediction task in Tencent's three online advertising scenarios: WeChat Channels, WeChat Moments, and the DSP.
% The backbone model employs Heterogeneous Experts with Multi-Embedding architecture~\cite{multi-embedding2023, stem2023, pan2024ad}.
% Specifically, we learn multiple different feature interaction experts, \textit{e.g.}, GwPFM~\cite{pan2024ad} (a variant of FFM~\cite{ffm2017} and FwFM~\cite{fwfm2018}), IPNN~\cite{pnn2016}, DCN V2~\cite{wang2021dcn}, or FlatDNN for sparse ID features.
% Multiple embedding tables are learned for all features, each corresponding to one or several experts.
% We employ a target-attention~\cite{din2018} model TIN~\cite{tin2023} for sequence features. 

We conduct an empirical evaluation of Combined-Pair on the Click-Through Rate (CTR) prediction task across three distinct online advertising scenarios in Tencent: \textit{WeChat Channels}, \textit{WeChat Moments}, and the \textit{Demand-Side Platform (DSP)}. 

\subsection{Deployment Details}
The underlying model architecture utilizes Heterogeneous Experts with Multi-Embedding framework~\cite{pan2024ad}.
Specifically, our approach involves training multiple feature interaction experts, such as GwPFM~\cite{pan2024ad} (a variant of FFM~\cite{ffm2017} and FwFM~\cite{fwfm2018}), IPNN~\cite{pnn2016}, DCN V2~\cite{wang2021dcn}, or FlatDNN, to capture diverse feature interactions for sparse ID features. 
Additionally, we learn multiple embedding tables for all features~\cite{me2023, stem2023, pan2018field}, with each table corresponding to one or several experts. 
For sequence features, we employ the TIN~\cite{tin2023} to capture the semantic-temporal correlation.

Based on the above backbone architecture, we deployed Combined-Pair and conducted online A/B testing from early July 2023 to August 2023. 
We configured the Combined-Pair with $\alpha=0.9$ and adopted streaming training.
The CTR varies from $0.1\%$ to $2.0\%$ in different scenarios.

The ranking loss adds an extra computation cost of $O(N_+\times N_-)$ per batch, where $N_+$ and $N_-$ denote the number of positive and negative samples within a batch. 
Such additional computation cost is negligible compared with the complexity of the backbone model. 
Besides, the ranking loss only influences the training time and doesn’t affect the inference time. 
During the online A/B test, the inference time and QPS are consistent with the baseline.

\subsection{Overall Performance}

\begin{figure}[!h]
    \centering
    \includegraphics[width=1\linewidth]{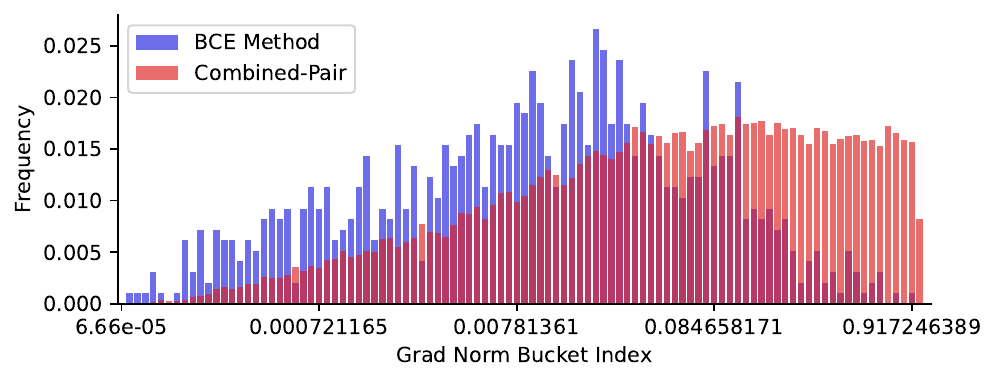}
    \caption{Distribution of gradient norms for negative samples in BCE method and Combined-Pair in online experiments.}
    \label{fig: online_grad_cmp}
\end{figure}

We examined the gradient distribution of negative samples and normalized their frequency, as shown in Fig.~\ref{fig: online_grad_cmp}. 
The analysis reveals that the gradient distribution of negative samples for the Combined-Pair method is significantly right-skewed compared to the BCE method, indicating that the Combined-Pair method obtains larger gradients.

Upon further examination of the model's online performance metrics, we observed that the Combined-Pair method significantly improves all business metrics compared to the BCE method, as shown in Tab.~\ref{tab: online_performance}. 
For instance, during one month of A/B testing with 20\% traffic, the Combined-Pair method achieves a cost lift of 0.59\% and a Gross Merchandise Value (GMV) lift of 0.70\% over the BCE method in the WeChat Moments scenario.

In addition, we also examined the \LossName{} for online deployment. 
Our observations indicate that the reduction in \LossName{} for the Combined-Pair method ranges from 0.01\% to 0.1\% over a span of 7 days compared to the BCE method during the A/B testing period.

\subsection{New Ads Performance}
Given that new ads have only a few training samples and are more prone to optimization difficulties, we specifically focused on the performance of the Combined-Pair method in the WeChat Channels scenario. 
The results, as shown in Tab.~\ref{tab: online_performance_new_ads}, demonstrate a significant performance improvement achieved by the Combined-Pair method. 
Specifically, it achieves a GMV lift of 1.26\% and a cost lift of 0.34\%, which are statistically significant, as confirmed by the \textit{t}-test.

\begin{table}[]
    \caption{Online A/B Testing Results.}
    \label{tab: online_performance}
    \begin{tabular}{@{}l|ccc@{}}
    \toprule
    \textbf{Ad Scenario} & \textbf{CTR}& \textbf{GMV} & \textbf{Cost}    \\ \midrule
    WeChat Channels & +0.91\% & +1.08\% & +0.29\% \\
    WeChat Moment   & +0.16\% & +0.70\% & +0.59\% \\
    DSP            & -0.04\% & +0.55\% & +0.15\% \\ \bottomrule
    \end{tabular}
\end{table}

\begin{table}[]
    \caption{Online A/B Testing Results for New Ads.}
    \label{tab: online_performance_new_ads}
    \begin{tabular}{@{}l|ccc@{}}
    \toprule
    \textbf{Launch Date} & \textbf{GMV} & \textbf{Cost}    \\ \midrule
    T    & +1.04\% & +0.27\% \\
    T-1    & +1.04\% & +0.27\% \\
    T-2    & +0.83\% & +0.47\% \\
    T-3             & +0.81\% & +0.17\% \\ \midrule
    Total  & +1.26\% & +0.34\% \\\bottomrule
    \end{tabular}
\end{table}

\section{Conclusion}

In this paper, we have identified a challenge associated with using only binary-cross entropy loss for Click-Through Rate (CTR) prediction when positive feedback is sparse. Specifically, we have observed the issue of gradient vanishing for negative samples in such scenarios.
To address this challenge, we propose a novel perspective by introducing an auxiliary ranking loss. 
We explain that the inclusion of this additional ranking loss leads to the generation of larger gradients for negative samples, effectively mitigating the problem of gradient vanishing.
Through comprehensive experiments and analysis, we have provided strong evidence to support our perspective. 
% We believe that this gradient-based perspective can significantly contribute to the optimization of CTR prediction models and serve as inspiration for further research in this field.
\begin{acks}
We want to express our sincere gratitude to the following individuals for their invaluable contributions to this research regarding the analysis, online implementation, and experiments: 
Ming Yue,
Kaixin Li,
Zhixiang Feng,
Xian Hu,
Yaqian Zhang,
Shifeng Wen,
Jiancheng Wang,
Yiding Deng,
Zeen Xu,
Xiaochen Wang,
Chen Cai,
GenBao Chen,
Chaonan Guo,
Junjie Zhai.
\end{acks}

\clearpage

\bibliographystyle{ACM-Reference-Format}
\balance
\bibliography{reference}

\end{document}